\documentclass[11pt]{article}

\usepackage[preprint]{acl}

\usepackage{times}
\usepackage{latexsym}
\usepackage{amsmath}
\usepackage{amssymb}

\usepackage{url}

\usepackage[T1]{fontenc}

\usepackage[utf8]{inputenc}

\usepackage{microtype}

\usepackage{inconsolata}

\usepackage{graphicx}
\usepackage{capt-of} 
\usepackage{booktabs}
\usepackage{multirow}
\usepackage{tabularx}
\usepackage{ragged2e}

\newcolumntype{F}{>{\ttfamily\small\raggedright\arraybackslash}p{0.42\textwidth}}
\newcolumntype{V}{>{\raggedright\arraybackslash}X}

\usepackage{listings}
\usepackage{xcolor}

\lstdefinelanguage{json}{
  basicstyle=\ttfamily\footnotesize,
  numbers=none,
  numberstyle=\scriptsize,
  stepnumber=1,
  numbersep=8pt,
  showstringspaces=false,
  breaklines=true,
  frame=lines,
  backgroundcolor=\color{white},
  literate=
   *{0}{{{\color{blue}0}}}{1}
    {1}{{{\color{blue}1}}}{1}
    {2}{{{\color{blue}2}}}{1}
    {3}{{{\color{blue}3}}}{1}
    {4}{{{\color{blue}4}}}{1}
    {5}{{{\color{blue}5}}}{1}
    {6}{{{\color{blue}6}}}{1}
    {7}{{{\color{blue}7}}}{1}
    {8}{{{\color{blue}8}}}{1}
    {9}{{{\color{blue}9}}}{1}
    {:}{{{\color{red}{:}}}}{1}
    {,}{{{\color{red}{,}}}}{1}
    {\{}{{{\color{violet}{\{}}}}{1}
    {\}}{{{\color{violet}{\}}}}}{1}
    {[}{{{\color{violet}{[}}}}{1}
    {]}{{{\color{violet}{]}}}}{1},
}

\title{Beyond Relevance: Structured Semantic Supervision for Product Search with LLM-Augmented Annotations}

\author{
  \textbf{Girish A.\ Koushik\textsuperscript{1}},
  \textbf{Swapnil Bhosale\textsuperscript{2}},
  \textbf{Samarth Agrawal\textsuperscript{3}},
  \textbf{Hadeel Sadany\textsuperscript{4}}\\
  \textbf{Constantin Orasan\textsuperscript{5}},
  \textbf{Xiatian Zhu\textsuperscript{2}},
  \textbf{Diptesh Kanojia\textsuperscript{2}}\\
  \textsuperscript{1}Computer Science, University of Surrey, Guildford, United Kingdom\\
  \textsuperscript{2}People-Centred AI, University of Surrey, Guildford, United Kingdom\\
  \textsuperscript{3}eBay USA\\
  \textsuperscript{4}Birmingham City University, Birmingham, United Kingdom\\
  \textsuperscript{5}Centre for Translation Studies, University of Surrey, Guildford, United Kingdom
}

\begin{document}
\maketitle
\begin{abstract}
E-commerce search requires distinguishing products that are merely related to a query from those that directly satisfy the user's shopping intent. We augment query-product pairs with structured LLM-generated query and product attributes and human-validated relevance, explanations, and centrality judgments, and evaluate these signals using a simple dual-encoder retriever and MLP re-ranker. On an augmented subset of ESCI, a human-feature oracle reaches $0.9382$ nDCG@10, while a human-free trained $Q+P$ configuration reaches $0.9258$. Synthetic approximations of the human signals reach $0.9150$ overall but provide substantial gains for difficult, low-performing queries. Ablations show that most of the oracle improvement comes from post-edited explanations and annotator comments rather than the scalar centrality feature, suggesting that LLMs are most useful for exposing and approximating structured semantic supervision rather than replacing human judgment directly.
\end{abstract}


\section{Introduction}  \label{sec:intro}

E-commerce search presents a unique retrieval challenge that differs fundamentally from traditional web search. The user queries tend to be very short, under-specified, and high in intent, requiring the mapping of a handful of keywords and/or phrases to millions of heterogeneous products. For example, the user might simply search for \textit{iphone}, \textit{nike shoes}, or \textit{memory foam mattress topper}, each implicitly incorporating brand considerations, category choices, compatibility considerations, and quality preferences that the retrieval system must infer. Previous research has consistently found that identifying the user intent, alternative product options, and purchasing behaviours are crucial for retrieving and ranking appropriate products, but this information is not easily captured by the queries themselves~\cite{ai2017learning,liu2018entity,manchanda2019intent}. As a result, e-commerce retrieval systems need to reason beyond lexical matching and identify products that best satisfy the underlying shopping intent.

A major challenge in product search is distinguishing products that are merely related to a query from those that directly satisfy the user's need: accessories, companion products, and loosely associated items may share semantic overlap with queries such as \textit{``iPad Pro case''} or \textit{``iPhone screen protector''} while still missing the core purchase intent~\cite{mcauley2015inferring}. Prior work has shown that modelling product semantics, user intent, and relevance granularity is critical in marketplace search, where ranking errors directly affect user satisfaction and conversion outcomes~\cite{chang2011detecting,guo2020debiasing,li2021embedding}. However, most retrieval systems still rely on coarse relevance labels that provide limited insight into why a product is relevant or how strongly it aligns with the primary intent. LLMs offer a way to expose such intermediate semantic signals by inferring latent attributes, generating explanations, and extracting structure from query-product pairs; nevertheless, because synthetic judgments may diverge from human relevance notions~\cite{faggioli2023perspectives,thomas2024large}, especially for fine-grained distinctions among substitutes, complements, and irrelevant products, we use LLMs to support human annotation and model training rather than replace human judgment.

In this work, we investigate whether structured supervision from LLM-generated attributes, validated through human annotation, can improve product retrieval and re-ranking. Starting from a subset of the Amazon ESCI benchmark~\cite{reddy2022shopping}, we augment each query-product pair with LLM-generated query attributes, product attributes, and relevance explanations; human annotators then refine these signals while assigning relevance labels and a binary centrality label~\cite{saadany2024centrality} indicating whether a product directly satisfies the primary query intent. We use this data to train a practical two-stage architecture comprising a lightweight dual-encoder retriever and an MLP-based re-ranker that operates on textual, categorical, and numerical features. Experiments show that structured intermediate supervision improves re-ranking beyond conventional relevance labels, while synthetic LLM-generated attributes achieve competitive performance without additional human annotation, especially for difficult, low-performing queries.



The contributions of this paper are as follows:
\begin{itemize}
    \item We introduce an LLM-augmented, human re-annotated product-search dataset containing structured query and product attributes, relevance explanations, re-graded relevance labels, and a binary centrality judgment that distinguishes intent-satisfying from peripheral products.
    \item We evaluate this structured supervision in a simple two-stage retrieval framework while explicitly separating a human-free setting from a human-feature oracle and an LLM-generated synthetic substitute, allowing us to distinguish deployable signals from diagnostic upper bounds.
    \item Through component ablations and query-level analysis, we show that human textual corrections drive most of the oracle gain, while centrality contributes little independently, and that synthetic annotations are most useful selectively for difficult queries rather than as a universal augmentation strategy.
\end{itemize}

\section{Related Work}    \label{sec:lit_review}

The Amazon Shopping Queries dataset, also known as ESCI, is the primary public benchmark for graded e-commerce query-product relevance~\cite{reddy2022shopping}. Its English Task-1 subset contains $29,844$ unique queries and $601,462$ judgments, including $20,888$ English training queries and $4,477$ public-test queries. Relevance is annotated using four ordered categories---Exact, Substitute, Complement, and Irrelevant, which are evaluated with gains of $1.0$, $0.1$, $0.01$, and $0.0$, respectively. Prior ESCI results show both the strength of interaction models and the difficulty of product relevance: a fine-tuned cross-encoder/MPNet setup substantially outperforms BM25, while the winning semantic-alignment system~\cite{zhang2022semantic} reaches $0.9043$ nDCG by training transformer classifiers over ESCI labels and converting class probabilities into ranking scores with augmentation, adversarial training, pseudo-labelling, self-distillation, and ensembling. Beyond ESCI, TREC Product Search 2023 adapts the setting to full-catalog retrieval and finds that lexical and sparse/hybrid systems remain highly competitive against general-purpose dense embeddings~\cite{campos2023overview}. Smaller benchmarks such as WANDS~\cite{chen2022wands} and Home Depot~\cite{home-depot-product-search-relevance} are useful for transfer evaluation, with \citet{chaudhary2023exploring} showing that direct transfer from ESCI outperforms label-conditioned synthetic query generation, reinforcing the value of in-domain labelled data.

Recent retrieval work has moved beyond single-vector dense models and cross-encoder re-ranking toward multi-vector, efficiency-aware, and LLM-assisted pipelines. Late-interaction models such as ColBERT preserve token-level query-document matching while allowing document representations to be precomputed, offering a practical middle ground between dual encoders and cross-encoders~\cite{khattab2020colbert}; later systems such as PLAID~\cite{santhanam2022plaid} and SLIM~\cite{li2023slim} further reduce serving costs through optimised centroid pruning and sparse inverted-index-compatible representations. In parallel, LLMs have been used in product search mainly for synthetic query/listing generation, automatic relevance labelling, and query rewriting: Aug2Search~\cite{xi2025aug2search} generates synthetic marketplace queries and enhanced listings for embedding-based retrieval, \citet{sachdev2024automated} uses LLMs for automated query-product relevance labelling, and \citet{ling2026synthetic} applies LLM-driven rewriting and synthetic query-product construction to long-tail retrieval. 


Our work sits at the intersection of these threads: we use an LLM as an annotation \emph{prior} to expose query and product attributes, along with relevance assessments and explanations, but do not treat its judgments as ground truth. Instead, human annotators post-edit these signals to produce final relevance labels and, before assigning relevance, provide a binary centrality judgment that captures whether the product directly satisfies the primary query intent. Unlike prior work that treats centrality primarily as a ranking signal~\citep{saadany2024centrality}, we study it as one component of a broader structured annotation framework and explicitly measure its marginal contribution alongside human textual explanations and comments.

\begin{figure*}[!ht]
\centering
\small
\includegraphics[width=0.8\textwidth]{"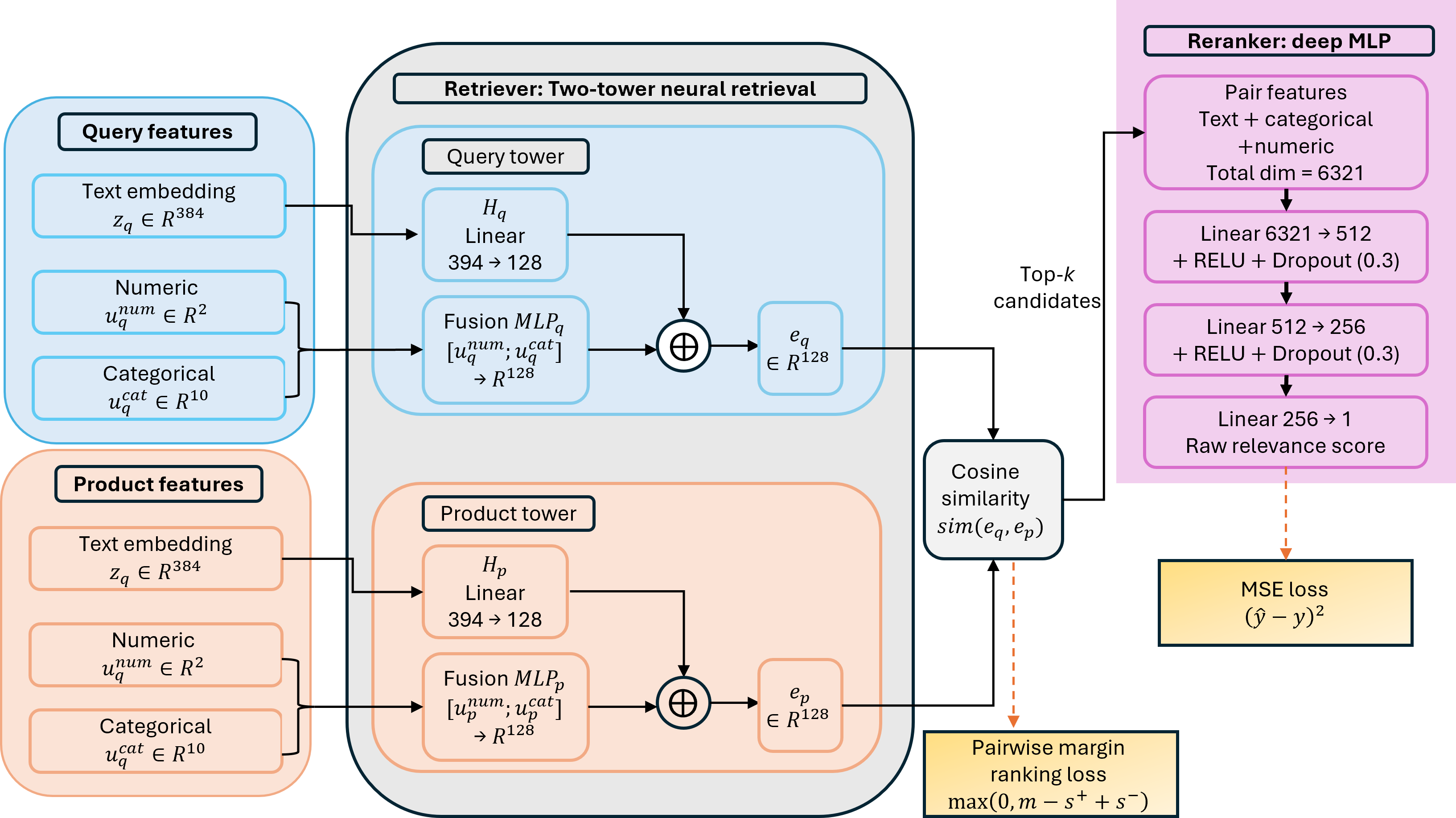"}
\caption{Overview of the proposed two-stage retrieval-reranker architecture.}
\label{fig:ebay-architecture}
\end{figure*}

\section{Dataset}   \label{sec:dataset}

We build the training and evaluation data on a sample of the ESCI dataset, augment each row with LLM-derived intermediate attributes, and re-annotate the augmented rows with human relevance, centrality score, and post-edit judgments. 

\begin{table}[!h]
\centering
\footnotesize
\setlength{\tabcolsep}{4pt}
\renewcommand{\arraystretch}{1.15}
\resizebox{0.9\columnwidth}{!}{%
\begin{tabular}{lrrr}
\toprule
Partition & Rows & unique (q,~p) & unique q \\
\midrule
Batch~1 (joint) & 120 & 120 & 86 \\
Batch~2 & 250 & 249 & 150 \\
Batch~3 & 250 & 250 & 155 \\
Batch~4 & 250 & 250 & 146 \\
Batch~5 & 250 & 249 & 156 \\
Batch~6 & 832 & 829 & 241 \\
\midrule
Annotated total & 1{,}952 & 1{,}887 & 266 \\
\midrule
Train split & --- & 903 & 191 \\
Test split  & --- & 984 &  75 \\
\bottomrule
\end{tabular}
}
\caption{Corpus statistics per batch. ``Rows'' is the number of (q,~p) rows; ``unique~(q,~p)'' deduplicates by surface form; ``unique~q'' is the number of distinct queries. Batch~1 was double-annotated by A1 and A2 (joint calibration); The rest were single-annotated.}
\vspace{-0.1cm}
\label{tab:dataset-stats}
\end{table}

\subsection{Source and Ordinal Label Mapping}

We sample $266$ unique queries from the English Task-1 subset of ESCI, giving $1{,}887$ unique (query, product title) pairs with $1{,}878$ distinct product titles. Original four-category ESCI labels were mapped to a four-point ordinal scale, mirroring the rank ordering. Each row carries median relevance and variance from the source data, summarising rater disagreement. 

\paragraph{LLM-based Intermediate Attributes.} Existing 4-point labels are sparse and do not explain \textit{why} a product is relevant or not. To surface a structured signal that downstream models can consume, we use Llama~3.1~8B-Instruct\footnote{\url{https://huggingface.co/meta-llama/Llama-3.1-8B-Instruct}} in zero-shot and few-shot settings to generate four query-side attributes and seven product-side attributes, each paired with a free-text explanation. For the query, we generate
\texttt{classification\_type}, \texttt{user\_intent}, \texttt{query\_complexity}, and \texttt{primary\_attribute\_focus}; for the product\_title we generate \texttt{product\_type}, \texttt{brand\_presence}, \texttt{title\_complexity},
\texttt{primary\_selling\_point}, \texttt{target\_demographic}, \texttt{price\_indicator}, and \texttt{condition} with set class labels. Further, the LLM is also asked to generate an explanation for each of these attributes. We finally ask the same model for a holistic \texttt{relevance\_assessment} label and an accompanying explanation that references the previously generated attributes. An example (LLM attributes plus human annotation) from our dataset is provided in Appendix~\ref{sec:appendix-data-sample}. Resulting columns form a structured intermediate representation used in three ways. First, human annotators read attribute labels and explanations before deciding the final centrality label and post-edited the relevance score. Second, the categorical attribute values are used as features for the retriever and re-ranker. Third, the post-edited relevance explanations serve as supervision targets for the synthetic-annotation generators (Section~\ref{sec:exp_results}).

\paragraph{Human Annotation.} As per guidelines provided, two annotators (A1 and A2) examine each row and \textbf{in this order}, 1) read the query, product title, and the LLM-generated intermediate attributes; 2) assign a binary \emph{centrality label} $c \in \{0,1\}$ reflecting whether the product is central to the query intent, distinct from surface relevance (a product can be ``relevant but peripheral'' \textit{e.g.}\ a phone case for the query ``iphone''); 3) assign a \emph{re-graded relevance score} on the same four-point ESCI-aligned scale, overriding the source label when they disagreed; 4) post-edit the LLM-generated relevance explanation, marking each edit as \texttt{no edit}, \texttt{minor edit}, or \texttt{major edit}, and optionally leaving a free-text comment. The two annotators jointly designed the rubrics for relevance, centrality, and post-edit categories. Annotation guidelines and full Inter-annotator reliability on the calibration set are in Appendix~\ref{sec:appendix-ann}.



\paragraph{Split and Data Statistics.} We use a strict \emph{query-level} split. All (query,~product) pairs sharing a query are placed in the same partition, eliminating the leakage that row-level splits introduce. The test partition contains $984$ pairs with $75$ unique queries, and is held out for all evaluation discussed in Section~\ref{sec:exp_results}. The remaining $191$ queries are used for training the retriever and re-ranker (Table~\ref{tab:dataset-stats}). On the $ 120$-query batch set, we retain the A1 ratings as the canonical label, given their experience with relevance grading.


\section{Two-Stage Architecture} \label{sec:arch}

We use a two-stage neural retrieval architecture designed for e-commerce search, operating through an initial retriever-based candidate generation phase followed by a precision-focused second-stage ranking process (shown in Figure~\ref{fig:ebay-architecture}). The foundation of the candidate generation phase is an asymmetric two-tower neural network where the query and product branches independently process distinct feature sets. For both towers, textual features are mapped into a $384$-dimensional space using a frozen pre-trained encoder~\cite{reimers2019sentence} (all-MiniLM-L6-v2\footnote{\href{https://huggingface.co/sentence-transformers/all-MiniLM-L6-v2}{Huggingface: transformers/all-MiniLM-L6-v2}} model). These text embeddings are projected into a $128$-dimensional shared latent space via a linear projection head. Concurrently, numeric scores and categorical attributes are fused through a multi-layer perceptron to form a $128$-dimensional dense auxiliary representation. The projected text embeddings are added to the fused auxiliary features to produce the final vectors $e_q$ and $e_p$. Candidate similarity is subsequently computed using the cosine similarity between the final query and product vectors, and the retriever is optimised using a standard Pairwise Margin Ranking Loss~\cite{karmaker2017application,paszke2019pytorch} function to enforce a minimum separation between positive and negative document scores.


\begin{table}[!ht]
\centering
\footnotesize
\renewcommand{\arraystretch}{1.15}
\resizebox{0.85\columnwidth}{!}{%
\begin{tabular}{lccc}
\toprule
Metric & MiniLM & Qwen3 & Diff. (pp) \\
\midrule
nDCG@3 & $\textbf{0.6051}$ & $0.5871$ & $-1.80$ \\
nDCG@5 & $\textbf{0.6468}$ & $0.6438$ & $-0.30$ \\
nDCG@10 & $0.7377$ & $\textbf{0.7441}$ & $+0.64$ \\
MRR@3 & $0.7935$ & $\textbf{0.8134}$ & $+1.99$ \\
MRR@5 & $0.7973$ & $\textbf{0.8172}$ & $+1.99$ \\
MRR@10 & $0.7973$ & $\textbf{0.8172}$ & $+1.99$ \\
Recall@3 & $\textbf{0.2528}$ & $0.2434$ & $-0.93$ \\
Recall@5 & $0.4178$ & $\textbf{0.4227}$ & $+0.49$ \\
Recall@10 & $0.7227$ & $\textbf{0.7431}$ & $+2.04$ \\
\bottomrule
\end{tabular}%
}
\caption{Embedding model comparison between MiniLM and Qwen3 using zero-shot cosine similarity (no training, pure embedding quality). Models Compared: (1) MiniLM: all-MiniLM-L6-v2 (384-dim, mean pooling, $22M$ params); (2) Qwen3-Embedding-0.6B (1024-dim, last token pooling, $600M$ params).}
\label{tab:embedding-model-comparison}
\vspace{-0.2cm}
\end{table}

\begin{table*}[!ht]
\centering
\footnotesize
\setlength{\tabcolsep}{7pt}
\renewcommand{\arraystretch}{1.15}
\resizebox{0.9\textwidth}{!}{%
\begin{tabular}{lccccccccc}
\toprule
\multirow{2}{*}{Approach} & \multicolumn{3}{c}{nDCG} & \multicolumn{3}{c}{MRR} & \multicolumn{3}{c}{Recall} \\
\cmidrule(lr){2-4}\cmidrule(lr){5-7}\cmidrule(lr){8-10}
 & @3 & @5 & @10 & @3 & @5 & @10 & @3 & @5 & @10 \\
\midrule
\midrule
\multirow{2}{*}{\shortstack[l]{MiniLM-L6 Cross-Encoder\\ (MS MARCO fine-tuned)}} & \multirow{2}{*}{\textbf{0.5724}} & \multirow{2}{*}{\textbf{0.5974}} & \multirow{2}{*}{\textbf{0.6770}} & \multirow{2}{*}{\textbf{0.8200}} & \multirow{2}{*}{\textbf{0.8200}} & \multirow{2}{*}{\textbf{0.8200}} & \multirow{2}{*}{\textbf{0.2894}} & \multirow{2}{*}{0.4275} & \multirow{2}{*}{\textbf{0.7051}} \\
 &  &  &  &  &  &  &  &  &  \\
\multirow{2}{*}{\shortstack[l]{e-Llama-3.1-8B\\~\cite{herold2025domain}}} & \multirow{2}{*}{0.5568} & \multirow{2}{*}{0.5913} & \multirow{2}{*}{0.6726} & \multirow{2}{*}{0.7356} & \multirow{2}{*}{0.7122} & \multirow{2}{*}{0.5367} & \multirow{2}{*}{0.2273} & \multirow{2}{*}{0.3478} & \multirow{2}{*}{0.3958} \\
 &  &  &  &  &  &  &  &  &  \\
MiniLM Bi-Encoder & 0.5406 & 0.5784 & 0.6591 & 0.7622 & 0.7656 & 0.7656 & 0.2792 & 0.4239 & 0.6989 \\
E5 Embeddings & 0.5192 & 0.5596 & 0.6354 & 0.7933 & 0.7933 & 0.7933 & 0.2864 & \textbf{0.4376} & 0.6983 \\
BM25 (Lexical) & 0.5077 & 0.5437 & 0.6285 & 0.7800 & 0.7800 & 0.7822 & 0.2758 & 0.4296 & 0.7005 \\
\bottomrule
\end{tabular}%
}
\caption{Traditional baselines on the original test set (984 Q-P pairs, 75 unique queries). Best scores are in bold.}
\label{tab:traditional-baselines}
\vspace{-0.1cm}
\end{table*}

Recognising that a fast bi-encoder sacrifices fine-grained contextual interaction, our architecture relies on a deep multi-layer perceptron re-ranker to rescue the final ranking quality for the top-$k$ candidates. The re-ranker ingests a $6321$-dimensional concatenated feature vector representing the query-product pair, comprising distinct text embeddings ($16*384$), categorical fields ($11*16$), and numeric scores ($1*1$). The network consists of three dense layers of decreasing size, transitioning from $6321$ to $512$ and then to $256$ dimensions, applying ReLU~\cite{nair2010rectified} activations and standard $0.3$ dropout at each hidden representation to mitigate memorisation. The final linear layer projects the hidden state to a single continuous scalar value representing the raw relevance score. Unlike the margin-based retriever, the re-ranker is trained as a regression task using a standard Mean Squared Error (MSE) loss~\cite{goodfellow2016deep} against ground-truth annotations, preserving absolute score calibration rather than relying strictly on relative ordering.


\paragraph{Feature availability and evaluation regimes.}
We distinguish three evaluation regimes throughout the experiments. $Q+P$ denotes the automatically generated query and product-side attributes and does not require human annotations at inference time. $H$ denotes human-provided signals: the post-edited relevance explanation, free-text annotator comment, and centrality label, and is therefore used only as an \emph{oracle diagnostic} to estimate the value of richer semantic supervision rather than as a deployable configuration. Finally, $\mathrm{Synth}\text{-}H$ replaces these human-only signals with LLM-generated surrogates and evaluates whether the information contained in $H$ can be approximated automatically. Accordingly, $Q+P$ is our human-free reference setting, $Q+P+H$ is an oracle upper bound, and $Q+P+\mathrm{Synth}\text{-}H$ tests the feasibility of automated approximation.

\section{Experiments and Results}   \label{sec:exp_results}

In this section, we present a comprehensive evaluation of our retrieval and re-ranking systems, using a strict query-level split across the dataset. 


\subsection{Pretrained and Zero-Shot Baselines}

As detailed in Table~\ref{tab:traditional-baselines}, the MS-MARCO Cross-Encoder~\cite{nogueira2019passage} delivers the strongest performance among traditional zero-shot baselines, achieving an nDCG@10 of $0.6770$ and a Recall@10 of $0.7051$. The generative e-Llama-3.1-8B~\cite{herold2025domain} model reaches a competitive nDCG@10 of $0.6726$ but a much lower Recall@10 of $0.3958$, indicating that it fails to retrieve a large portion of relevant items; its MRR also degrades from $0.7356$ at rank~3 to $0.5367$ at rank~10. This comparison is indicative rather than like-for-like, as e-Llama is a generative scorer evaluated here on global ranking. In contrast, the standard embedding baselines such as MiniLM~\cite{wang2020minilm} and E5~\cite{wang2022text} maintain stable MRR scores and consistent recall near $0.70$. Additionally, we experiment with ColBERT baselines in Appendix~\ref{sec:appendix-baseline}.

Table~\ref{tab:embedding-model-comparison} directly compares the pure embedding quality of MiniLM and Qwen3. Qwen3~\cite{yang2025qwen3} provides a marginal advantage in broad retrieval, scoring higher in nDCG@10 ($0.7441$ versus $0.7377$) and Recall@10 ($0.7431$ versus $0.7227$). MiniLM counters with stronger early precision, beating Qwen3 in nDCG@3 ($0.6051$ versus $0.5871$). Because the overall performance gap is negligible, MiniLM remains the practical baseline choice given faster inference and fewer parameters. These comparisons are intended to establish pretrained and zero-shot reference points on our evaluation split rather than to claim state-of-the-art performance against fully fine-tuned re-rankers. We do not fine-tune a cross-encoder on the re-annotated split; published ESCI systems that do so are discussed in Section~\ref{sec:lit_review}, but their reported scores are not directly comparable because they use different training and evaluation partitions.

\begin{table*}[!ht]
\centering
\footnotesize
\setlength{\tabcolsep}{7pt}
\renewcommand{\arraystretch}{1.15}
\resizebox{0.9\textwidth}{!}{%
\begin{tabular}{lccccccccc}
\toprule
\multirow{2}{*}{Configuration} & \multicolumn{3}{c}{nDCG} & \multicolumn{3}{c}{MRR} & \multicolumn{3}{c}{Recall} \\
\cmidrule(lr){2-4}\cmidrule(lr){5-7}\cmidrule(lr){8-10}
 & @3 & @5 & @10 & @3 & @5 & @10 & @3 & @5 & @10 \\
\midrule
Q Only (Query Attrs) & $\textbf{0.6334}$ & $\textbf{0.6640}$ & $\textbf{0.7305}$ & $\textbf{0.8358}$ & $\textbf{0.8395}$ & $\textbf{0.8395}$ & $\textbf{0.2895}$ & $\textbf{0.4508}$ & $\textbf{0.7371}$ \\
P Only (Product Attrs) & $0.6015$ & $0.6362$ & $0.7036$ & $0.8209$ & $0.8247$ & $0.8247$ & $0.2735$ & $0.4353$ & $0.7226$ \\
Q+P (Full Features) & $0.5942$ & $0.6250$ & $0.6939$ & $0.8308$ & $0.8346$ & $0.8346$ & $0.2703$ & $0.4276$ & $0.7186$ \\
\bottomrule
\end{tabular}%
}
\caption{Trained retriever-only results across feature configurations. Best scores are in bold.}
\label{tab:trained-retriever-only}
\end{table*}

\begin{table*}[!ht]
\centering
\footnotesize
\setlength{\tabcolsep}{7pt}
\renewcommand{\arraystretch}{1.15}
\resizebox{0.9\textwidth}{!}{%
\begin{tabular}{lllccccc}
\toprule
\multirow{2}{*}{Retriever} & \multirow{2}{*}{Configuration} & \multirow{2}{*}{Generator} & \multicolumn{3}{c}{nDCG} & MRR & Recall \\
\cmidrule(lr){4-6}\cmidrule(lr){7-7}\cmidrule(lr){8-8}
 &  &  & @3 & @5 & @10 & @10 & @10 \\
\midrule
Pretrained & Q+P+H & Human & $\mathbf{0.8880}$ & $\mathbf{0.9129}$ & $\mathbf{0.9382}$ & $\mathbf{1.0000}$ & $0.8150$ \\
Trained & Q+P+H & Human & $0.8845$ & $0.9074$ & $0.9382$ & $1.0000$ & $\mathbf{0.8156}$ \\
Trained & Q+H & Human & $0.8716$ & $0.9012$ & $0.9225$ & $0.9925$ & $0.8132$ \\
Trained & Q+P & None & $0.8523$ & $0.8869$ & $0.9258$ & $0.9826$ & $0.8016$ \\
Trained & Q+P+Synth-H & Qwen3.5-27B & $0.8821$ & $0.8820$ & $0.9150$ & $0.9701$ & $0.8034$ \\
Trained & Q+P+Synth-H & Mistral-7B & $0.7986$ & $0.8276$ & $0.8773$ & $0.9751$ & $0.7996$ \\
Pretrained & Q+P & None & $0.6968$ & $0.7404$ & $0.8089$ & $0.9321$ & $0.7679$ \\
\bottomrule
\end{tabular}%
}
\caption{Unified two-stage and annotation results against $75$ unique queries test set. Best scores are in bold.}
\label{tab:unified-two-stage-annotation}
\vspace{-0.1cm}
\end{table*}

\subsection{Model Performance and Validation}

We first isolate retriever-only performance across feature configurations. As shown in Table~\ref{tab:trained-retriever-only}, query attributes alone perform best, reaching $0.7305$ nDCG@10, while product-only attributes reach $0.7036$, and combining query and product attributes lowers performance to $0.6939$, suggesting that first-stage retrieval is sensitive to explicit feature combinations and that additional attributes can sometimes introduce noise. We then evaluate synthetic annotations as a scalable alternative to costly human labelling, targeting niche, multi-attribute, or brand-specific queries where the baseline struggles. Using Qwen3.5-27B\footnote{\url{https://huggingface.co/Qwen/Qwen3.5-27B}} and Mistral-7B\footnote{\url{https://huggingface.co/mistralai/Mistral-7B-Instruct-v0.3}} with few-shot prompting~\cite{brown2020language}, we generate relevance score, relevance explanations, and centrality score similar to the human-observed annotations in our dataset (see Appendix~\ref{sec:appendix-synth-ann} for an example), treating these synthetic features as a situational re-ranking fallback rather than a universal replacement for standard retrieval features. Direct agreement between the synthetic and human annotations is reported in Appendix~\ref{subsec:synthetic-human-agreement}; we therefore treat synthetic labels as approximate surrogates rather than interchangeable replacements for human judgments.

Table~\ref{tab:unified-two-stage-annotation} presents the unified results across the three feature-availability regimes described above. The human-feature oracle ($Q+P+H$) reaches an nDCG@10 of $0.9382$, quantifying the potential value of richer human semantic signals. Importantly, the trained human-free $Q+P$ configuration remains close at $0.9258$ nDCG@10 and requires no human-provided features at inference time. Replacing $H$ with Qwen3.5-27B-generated synthetic annotations yields $0.9150$ nDCG@10, while Mistral-7B reaches $0.8773$. Thus, synthetic $H$ does not improve aggregate performance over the trained $Q+P$ setting; its value is instead query-dependent, as examined in \S\ref{subsec:synthetic-analysis}.

To identify which human signals account for the oracle improvement, we perform a leave-one-out ablation over the three $H$ components (Appendix~\ref{sec:appendix-ablation}). Removing the post-edited relevance explanation reduces nDCG@10 by $0.0775$, and removing annotator comments reduces it by $0.0640$, whereas removing the binary centrality feature changes nDCG@10 by only $0.0010$. These results show that the principal gain from $H$ comes from richer textual supervision that corrects or sharpens the model-generated semantic interpretation. Centrality remains useful as an interpretable annotation dimension for distinguishing intent-satisfying from peripheral products, but it is not, by itself, the dominant predictive feature in the re-ranker. Finally, we report an additional validation on an internal test set using delta values relative to the \textit{Pretrained Retriever} baseline in Appendix~\ref{sec:appendix-ext}, Table~\ref{tab:spot-eval}.


\paragraph{Pre-trained vs. Trained Retriever.}
With human annotations, the re-ranker's strong supervision makes retriever quality largely irrelevant at the top of the ranking: pretrained and trained retrievers achieve identical nDCG@10 ($0.9382$) with Q+P+H. The pretrained retriever is slightly better at stricter cutoffs (nDCG@3: $0.8880$ vs. $0.8845$; nDCG@5: $0.9129$ vs. $0.9074$), suggesting that fine-tuning may over-specialise to training queries and narrow the pretrained embeddings' broader semantic coverage. Without human annotations, however, retriever training is crucial: the gap between pretrained and trained Q+P is $11.69$ pp (nDCG@10: $0.8089$ vs. $0.9258$). This asymmetry exposes a cost-efficiency tradeoff: annotation effort and retriever training can substitute for each other. Under annotation constraints, practitioners should prioritise retriever training; with ample annotations, they can skip retriever training without losing top-10 ranking quality.


\subsection{Synthetic Annotation Analysis}
\label{subsec:synthetic-analysis}

To understand when synthetic annotations help, we compare the standard Q+P baseline against Q+P+Synth-H at the query level. Table~\ref{tab:top-improve-synth} in Appendix~\ref{sec:appendix-qual} shows that the largest gains occur for difficult low-baseline queries: ``nike shoe'' improves from $0.5157$ to $0.9988$ nDCG@10 ($+0.4831$), with similar gains for ``golf shoes'' ($+0.4436$) and ``memory foam mattress topper'' ($+0.3233$), suggesting that LLM-generated semantic expansion can bridge vocabulary gaps missed by standard retrieval features. In contrast, Table~\ref{tab:top-degrade-synth} shows that synthetic context can hurt queries that already perform well: ``light bulb'' drops from a perfect $1.0000$ nDCG@10 to $0.9676$, and queries such as ``levis'' and ``polaroid camera'' show similar minor degradation. Taken together with the aggregate result in Table~\ref{tab:unified-two-stage-annotation}, this analysis does not support applying synthetic annotations universally. Instead, it motivates a gated strategy in which synthetic semantic expansion is invoked only for queries whose baseline ranking is uncertain or poor, with the human-free $Q+P$ model retained as the default path.

\section{Conclusion}    \label{sec:conclusion}

We presented a structured semantic supervision framework for product search that combines LLM-generated query and product attributes with human re-annotation of relevance, explanations, and centrality labels. Using a simple two-stage retriever and re-ranker, we separated a human-free $Q+P$ setting from a human-feature oracle and an automated synthetic score. The human-feature oracle achieves the strongest ranking quality, but the human-free trained model remains close, while synthetic annotations do not improve aggregate performance over this baseline. Their benefit is instead concentrated on difficult, low-performing queries, suggesting selective rather than universal use. Ablations further show that post-edited explanations and annotator comments account for most of the oracle gain, whereas the scalar centrality feature contributes little independently. Overall, the results support using LLMs to expose and approximate richer semantic supervision, while retaining human judgment as the reference rather than assuming synthetic labels are direct substitutes.

\section*{Limitations}

First, our human-annotated dataset is a subset of the ESCI benchmark, and we use a strict query-level split. The scale and domain coverage are still limited compared to full production e-commerce search. Second, our LLM-generated attributes and synthetic annotations are sensitive to the prompting setup and the models employed, and their reliability may differ across product categories, languages, marketplaces, and long-tail queries. Third, although synthetic augmentation is competitive on some queries, it does not improve aggregate performance over the human-free $Q+P$ setting; our qualitative findings suggest it can introduce noise into queries that previously performed well, implying its use will require proper query routing or confidence estimation. Fourth, the re-ranker concatenates several embeddings and the synthetic path uses a generative model, so deployment carries a latency and cost overhead that we do not quantify here. Fifth, our ablations isolate the human-provided annotation components, but do not fully disentangle all interactions among re-graded relevance labels, generated query/product attributes, and model architecture; consequently, we interpret the results as evidence for the structured-supervision pipeline as a whole rather than attributing all gains to any one signal. Finally, the external evaluation of the internal e-commerce corpus cannot be quantified with absolute metrics, and the consistent presence of certain ranking features suggests that this benchmark may not adequately assess the challenges in retrieval. It is therefore necessary that future research focus on evaluating the structured supervision paradigm on bigger, more complex product search data corpora. 



\bibliography{custom}

\clearpage

\onecolumn
\appendix

\section{Sample Annotation from our Dataset}
\label{sec:appendix-data-sample}

\begin{table*}[!ht]
\centering

\renewcommand{\arraystretch}{1.10}

\begin{tabularx}{\textwidth}{@{}FV@{}}
\toprule
\textbf{Column / Feature Field} &
\textbf{Dataset Value / Feature Content} \\
\midrule

query (q) &
minecraft diamond sword \\

product\_title (p) &
The Quest for the Diamond Sword: A Minecraft Gamer's Adventure \\


\midrule

\multicolumn{2}{@{}l@{}}{\small\bfseries LLM-Generated Query Categorical Fields} \\
\midrule

q\_expansion\_classification\_type\_value &
Feature-based \\

q\_expansion\_user\_intent\_value &
Research \\

q\_expansion\_query\_complexity\_value &
Simple keyword \\

q\_expansion\_primary\_attribute\_focus\_value &
Feature \\

\midrule

\multicolumn{2}{@{}l@{}}{\small\bfseries LLM-Generated Product Categorical Fields} \\
\midrule

p\_expansion\_product\_type\_value &
Specific Product \\

p\_expansion\_brand\_presence\_value &
Not Applicable \\

p\_expansion\_title\_complexity\_value &
Simple \\

p\_expansion\_primary\_selling\_point\_value &
Feature \\

p\_expansion\_target\_demographic\_value &
Specific (Minecraft Gamers) \\

p\_expansion\_price\_indicator\_value &
Not Indicated \\

p\_expansion\_condition\_value &
New \\

\midrule

\multicolumn{2}{@{}l@{}}{\small\bfseries LLM-Generated Explanation} \\
\midrule

relevance\_assessment\_result\_relevance &
Highly Relevant \\

relevance\_assessment\_result\_explanation &
The query and product title have a high degree of relevance, with exact matches on feature-based classification type, primary selling point, condition, and target demographic. \\

\midrule

\multicolumn{2}{@{}l@{}}{\small\bfseries Human Evaluation Labels} \\
\midrule

relevance\_score & 1.0 \\

centrality\_score &
1.0 \\

comments &
The query and product title have matches on feature-based classification type, primary selling point and condition. The query is for a Minecraft sword and the product is a book about the adventure, introducing a subtle target mismatch. \\

edit\_type &
Major edit \\

\bottomrule
\end{tabularx}
\caption{Dataset instance for the query ``minecraft diamond sword''.}
\label{tab:minecraft_dataset_sample}
\end{table*}

\section{ColBERT Baselines}
\label{sec:appendix-baseline}

Table~\ref{tab:colbert-late-interaction} outlines the zero-shot retrieval capabilities of ColBERT late-interaction models. The ColBERT-v1~\cite{khattab2020colbert} Vanilla configuration strictly utilizing query and product text outperforms all traditional baselines, reaching an nDCG@10 of $0.7434$. Integrating explicit LLM-derived attributes degrades ColBERT's performance. Adding query attributes lowers the nDCG@10 to $0.7093$, and including both query and product attributes drops it further to 0.7028. ColBERT-v2~\cite{santhanam2022colbertv2} Vanilla provides slightly lower absolute performance than v1, with an nDCG@10 of $0.7258$.

\begin{table*}[!ht]
\centering
\footnotesize
\setlength{\tabcolsep}{6pt}
\renewcommand{\arraystretch}{1.15}
\resizebox{\textwidth}{!}{%
\begin{tabular}{llccccccccc}
\toprule
\multirow{2}{*}{Model} & \multirow{2}{*}{Configuration} & \multicolumn{3}{c}{nDCG} & \multicolumn{3}{c}{MRR} & \multicolumn{3}{c}{Recall} \\
\cmidrule(lr){3-5}\cmidrule(lr){6-8}\cmidrule(lr){9-11}
 &  & @3 & @5 & @10 & @3 & @5 & @10 & @3 & @5 & @10 \\
\midrule
ColBERT v1 & Vanilla (Q vs P) & \textbf{0.6245} & \textbf{0.6657} & \textbf{0.7434} & 0.8209 & 0.8276 & 0.8276 & 0.2603 & 0.4309 & 0.7247 \\
ColBERT v2 & Vanilla (Q vs P) & 0.5863 & 0.6411 & 0.7258 & 0.7861 & 0.7898 & 0.7898 & 0.2387 & 0.4228 & 0.7202 \\
ColBERT v1 & Q Only & 0.5943 & 0.6288 & 0.7093 & 0.8358 & 0.8396 & 0.8396 & \textbf{0.2829} & 0.4434 & \textbf{0.7325} \\
ColBERT v1 & Q+P & 0.5862 & 0.6192 & 0.7028 & \textbf{0.8458} & \textbf{0.8495} & \textbf{0.8495} & 0.2742 & \textbf{0.4449} & 0.7253 \\
ColBERT v1 & Q+P+H & 0.5590 & 0.5958 & 0.6956 & 0.8308 & 0.8368 & 0.8368 & 0.2738 & 0.4255 & 0.7265 \\
ColBERT v1 & Q+H & 0.5478 & 0.5861 & 0.6872 & 0.8209 & 0.8276 & 0.8301 & 0.2664 & 0.4178 & 0.7230 \\
\bottomrule
\end{tabular}%
}
\caption{ColBERT late-interaction results with zero-shot retrieval (pre-trained, no domain-specific training) and LLM-derived attributes. Configurations: Vanilla (Q+P text only), Q (query attributes), Q+P (query + product attributes), Q+H, Q+P+H. Best scores are in bold.}
\label{tab:colbert-late-interaction}
\end{table*}

\section{External Validation}
\label{sec:appendix-ext}

The validation on the internal test set is reported using delta values relative to the \textit{Pretrained Retriever} baseline, rather than absolute metrics. The strongest relative gains are observed for ColBERT, consistently outperforming the baseline on nDCG and Recall across the evaluated cutoffs. The Trained Retriever and Trained Reranker also improve over the baseline, although by smaller margins than the ColBERT models. In contrast, e-Llama-3.1-8B shows positive relative gains for nDCG and early Recall, but a substantial degradation in MRR@10 and Recall@10, suggesting that its independent prompt-based scoring is less reliable for global ranking over the full candidate set.

\begin{table*}[!ht]
\centering
\footnotesize
\setlength{\tabcolsep}{6pt}
\renewcommand{\arraystretch}{1.15}
\resizebox{\textwidth}{!}{%
\begin{tabular}{lccccccccc}
\toprule
\multirow{2}{*}{Model} & \multicolumn{3}{c}{$\Delta$nDCG} & \multicolumn{3}{c}{$\Delta$MRR} & \multicolumn{3}{c}{$\Delta$Recall} \\
\cmidrule(lr){2-4}\cmidrule(lr){5-7}\cmidrule(lr){8-10}
 & @3 & @5 & @10 & @3 & @5 & @10 & @3 & @5 & @10 \\
\midrule
\shortstack[l]{ColBERT v2\\(Vanilla)} & $\mathbf{+0.0593}$ & $\mathbf{+0.0623}$ & $\mathbf{+0.0550}$ & $+0.0000$ & $+0.0000$ & $+0.0000$ & $\mathbf{+0.1825}$ & $\mathbf{+0.2974}$ & $\mathbf{+0.2662}$ \\
\shortstack[l]{ColBERT v1\\(Vanilla)} & $+0.0590$ & $+0.0621$ & $+0.0547$ & $+0.0000$ & $+0.0000$ & $+0.0000$ & $+0.1825$ & $+0.2974$ & $+0.2662$ \\
e-Llama-3.1-8B & $+0.0512$ & $+0.0557$ & $+0.0495$ & $-0.0069$ & $-0.0203$ & $-0.4771$ & $+0.1777$ & $+0.2789$ & $-0.2105$ \\
\shortstack[l]{Trained\\Retriever} & $+0.0368$ & $+0.0451$ & $+0.0442$ & $+0.0000$ & $+0.0000$ & $+0.0000$ & $+0.1825$ & $+0.2974$ & $+0.2662$ \\
\shortstack[l]{Trained\\Reranker} & $+0.0189$ & $+0.0323$ & $+0.0360$ & $+0.0000$ & $+0.0000$ & $+0.0000$ & $+0.1825$ & $+0.2974$ & $+0.2662$ \\
\shortstack[l]{Pretrained\\Retriever} & $+0.0000$ & $+0.0000$ & $+0.0000$ & $+0.0000$ & $+0.0000$ & $+0.0000$ & $+0.0000$ & $+0.0000$ & $+0.0000$ \\
\bottomrule
\end{tabular}%
}
\caption{External validation on an internal test set. To protect internal absolute metrics, all values are reported as deltas relative to the \textit{Pre-trained Retriever} baseline. The baseline row is shown as $0.0000$. Configuration: Text-only query vs product title, no LLM attributes, with shuffling. Best scores are in bold.}
\label{tab:spot-eval}
\end{table*}

Despite the relative improvements on the internal test set, the results should be interpreted cautiously. Several embedding-based models exhibit identical relative Recall gains and no observable MRR differences from the baseline, suggesting that the benchmark may not sufficiently distinguish strong ranking systems in this evaluation setup. Since the dataset was shuffled within queries to reduce positional bias, the uniformity of the ranking metrics likely reflects limitations in benchmark difficulty rather than conclusive evidence of production robustness. These results are therefore best treated as relative upper-bound validation signals on an internal dataset, not as direct estimates of real-world search performance.

\section{Annotation Analysis}
\label{sec:appendix-ann}

\paragraph{Annotation Design and Agreement Protocol.}
The corpus was labelled by two annotators. A1 annotated all $1{,}887$ pairs and provides the canonical labels used throughout. Before the guidelines were frozen, A2 independently re-annotated the first batch of $120$ pairs as a calibration set, so that we could align the rubric and quantify inter-annotator reliability; the remaining pairs are single-annotated by design, with A1 as the canonical rater. We report agreement on this $120$-pair calibration set, and note that a further $13$ pairs recurring across later batches were also judged by both annotators and give consistent estimates. Because relevance is an ordered four-point scale, on which nominal agreement understates reliability, we report raw exact agreement, within-one-point agreement, and quadratic-weighted Cohen's $\kappa$, which penalises distant disagreements more than adjacent ones. Centrality is binary, so we report unweighted Cohen's $\kappa$ alongside exact agreement. Centrality is also skewed, and the two annotators' marginals differ, so we also report prevalence- and bias-robust coefficients that are not deflated by the high-agreement $\kappa$ paradoxes.

\begin{table}[!ht]
\centering
\begin{minipage}[t]{0.48\textwidth}
\centering
\small
\setlength{\tabcolsep}{4pt}
\renewcommand{\arraystretch}{1.15}
\begin{tabular}{lc}
\toprule
Quantity & Count \\
\midrule
Corpus labelled by A1 (canonical)        & $1{,}887$ \\
Calibration batch, double-annotated      & $120$ \\
Additional doubly judged cross-batch pairs & $13$ \\
Reliability set (doubly judged)          & $133$ \\
\bottomrule
\end{tabular}
\captionof{table}{Annotation coverage and the doubly judged sets used for reliability.}
\label{tab:iaa-design}
\end{minipage}\hfill
\begin{minipage}[t]{0.48\textwidth}
\centering
\small
\setlength{\tabcolsep}{8pt}
\renewcommand{\arraystretch}{1.15}
\begin{tabular}{lcccc}
\toprule
A1 $\backslash$ A2 & 1 & 2 & 3 & 4 \\
\midrule
1 & $31$ & $16$ & $6$  & $0$ \\
2 & $7$  & $9$  & $16$ & $0$ \\
3 & $1$  & $5$  & $23$ & $6$ \\
4 & $0$  & $0$  & $0$  & $0$ \\
\bottomrule
\end{tabular}
\captionof{table}{Relevance confusion matrix on the calibration batch ($n{=}120$); mass concentrates on and adjacent to the diagonal.}
\label{tab:conf-rel}
\end{minipage}
\end{table}

\subsection{Choice of Coefficients}
Single-coefficient reporting is unsafe here for two reasons: relevance is an ordered four-point scale, so nominal agreement understates reliability; and centrality is a skewed binary label with unequal annotator marginals, the regime in which Cohen's $\kappa$ is depressed by the prevalence and bias paradoxes~\cite{cohen1960coefficient,cohen1968weighted,landis1977measurement,gwet2008computing,cicchetti1990high,krippendorff2004content,artstein2008inter}. We therefore report: raw and within-one agreement; chance-corrected Cohen's $\kappa$ with linear and quadratic weights for the ordinal relevance; Scott's $\pi$ and Krippendorff's $\alpha$ (nominal and ordinal), which assume a common marginal and handle the ordinal structure; Gwet's AC1 and the prevalence-adjusted bias-adjusted $\kappa$ (PABAK), which are stable under skew; rank association (Spearman $\rho$, Pearson $r$, Kendall $\tau_b$); and, for centrality, a McNemar test of marginal symmetry with prevalence and bias indices. 

\subsection{Relevance}
Relevance agreement is moderate to substantial once the ordinal structure is respected (Table~\ref{tab:iaa-rel}). Although exact agreement is $52.5\%$ and unweighted $\kappa$ is $0.30$, $94.2\%$ of all judgments agree within a single point, the quadratic-weighted $\kappa$ is $0.61$, the ordinal Krippendorff $\alpha$ is $0.61$, and rank association is $\rho = 0.65$. The confusion matrix (Table~\ref{tab:conf-rel}) shows that disagreements are almost entirely between adjacent grades (Exact vs Substitute, Substitute vs Complement) rather than categorical confusions, which is the expected profile for graded e-commerce relevance and is consistent with the low unweighted but substantial weighted coefficients.

\subsection{Centrality}
Centrality is the harder judgment and shows moderate agreement that is governed by a difference in annotator stringency rather than random noise (Table~\ref{tab:iaa-cen}). Exact agreement is $68.3\%$ with Cohen's $\kappa = 0.42$, but A1 marks $69.2\%$ of items central against $39.2\%$ for A2, a marginal asymmetry that a McNemar test rejects as chance ($\chi^2 = 32.2$, $p < 10^{-7}$). This disagreement is one-sided. Of the $37$ items, A1 judged not central, A2 agreed on $36$; the gap lies entirely in the central direction (bias index $0.30$). Because Cohen's $\kappa$ is suppressed under this combination of high prevalence and marginal bias, we also report Gwet's AC1 ($0.37$) and PABAK ($0.37$), and the positive and negative specific-agreement values ($0.71$ and $0.66$), which bracket the chance-corrected estimates. We adopt A1, who labelled every batch, as the canonical annotator so that the released centrality labels are internally consistent and not contaminated by inter-annotator stringency drift; the guidelines were frozen after this calibration batch.

\begin{table}[!ht]
\centering
\begin{minipage}[t]{0.48\textwidth}
\centering
\small
\setlength{\tabcolsep}{6pt}
\renewcommand{\arraystretch}{1.15}
\begin{tabular}{lc}
\toprule
Coefficient & Calibration ($n{=}120$) \\
\midrule
Exact agreement             & $52.5\%$ \\
Within-one agreement        & $94.2\%$ \\
Mean absolute error         & $0.53$ \\
Cohen's $\kappa$            & $0.30$ \\
Linear-weighted $\kappa$    & $0.47$ \\
Quadratic-weighted $\kappa$ & $0.61$ \\
Scott's $\pi$               & $0.30$ \\
Krippendorff $\alpha$ (ord.)& $0.61$ \\
Gwet's AC1                  & $0.39$ \\
PABAK                       & $0.37$ \\
Spearman $\rho$             & $0.65$ \\
Kendall $\tau_b$            & $0.59$ \\
\bottomrule
\end{tabular}
\captionof{table}{Relevance reliability on the $120$-pair calibration set (four-point ordinal scale).}
\label{tab:iaa-rel}
\end{minipage}\hfill
\begin{minipage}[t]{0.48\textwidth}
\centering
\small
\setlength{\tabcolsep}{6pt}
\renewcommand{\arraystretch}{1.15}
\begin{tabular}{lc}
\toprule
Coefficient & Calibration ($n{=}120$) \\
\midrule
Exact agreement     & $68.3\%$ \\
Cohen's $\kappa$    & $0.42$ \\
Scott's $\pi$       & $0.36$ \\
Krippendorff $\alpha$ & $0.37$ \\
Gwet's AC1          & $0.37$ \\
PABAK               & $0.37$ \\
Positive agreement  & $0.71$ \\
Negative agreement  & $0.66$ \\
Prevalence index    & $0.08$ \\
Bias index          & $0.30$ \\
A1 / A2 central     & $69.2\% / 39.2\%$ \\
McNemar $\chi^2$ ($p$) & $32.2\ (p{<}10^{-7})$ \\
\bottomrule
\end{tabular}
\captionof{table}{Centrality reliability on the $120$-pair calibration set (binary). Low chance-corrected values reflect annotator stringency (bias), so AC1 and PABAK are also reported.}
\label{tab:iaa-cen}
\end{minipage}
\end{table}

\begin{table}[!ht]
\centering
\begin{minipage}[t]{0.48\textwidth}
\centering
\footnotesize
\setlength{\tabcolsep}{6pt}
\renewcommand{\arraystretch}{1.15}
\begin{tabular}{lc}
\toprule
Change vs ESCI source label & Count (\%) \\
\midrule
Unchanged              & $726$ ($38.6\%$) \\
Decreased by one       & $763$ ($40.5\%$) \\
Increased by one       & $150$ ($8.0\%$) \\
Changed by two or more & $243$ ($12.9\%$) \\
\midrule
Any change             & $1{,}156$ ($61.4\%$) \\
\bottomrule
\end{tabular}
\captionof{table}{Human re-grading relative to the ESCI source label over the $1{,}887$ canonical pairs.}
\label{tab:relabel}
\end{minipage}\hfill
\begin{minipage}[t]{0.48\textwidth}
\centering
\footnotesize
\setlength{\tabcolsep}{5pt}
\renewcommand{\arraystretch}{1.15}
\begin{tabular}{lcc}
\toprule
 & Source & Re-graded \\
\midrule
Irrelevant (1) & $314$ & $673$ \\
Complement (2) & $558$ & $467$ \\
Substitute (3) & $509$ & $715$ \\
Exact (4)      & $501$ & $27$ \\
\midrule
Central (1)    & ---   & $1{,}422$ \\
Peripheral (0) & ---   & $458$ \\
\midrule
Post-edit no/min/maj & --- & $324/671/950$ \\
\bottomrule
\end{tabular}
\captionof{table}{Label distribution over the $1{,}887$ canonical pairs (centrality labelled on $1{,}880$; central $=75.6\%$; mean source variance $0.56$).}
\label{tab:label-dist}
\end{minipage}
\end{table}

\subsection{Post-edit Category}
Agreement on the three-way post-edit category (no/minor/major) is low (exact $45.2\%$, Cohen's $\kappa = 0.12$, Gwet's AC1 $0.21$, nominal $\alpha = 0.10$). Annotators agree on whether an edit occurred but differ on the no/minor boundary (A2 records ``no edit'' for $35.7\%$ of items against $14.8\%$ for A1). The post-edit category is an effort side signal used to study where supervision helps; it is not a label consumed by the model, so its lower reliability does not affect the released relevance and centrality labels.

\subsection{Re-annotation Adds Signal}
Re-annotation is not a rubber-stamp of the source labels. Relative to the ESCI source grade, the canonical annotator changed $61.4\%$ of labels ($40.5\%$ down one level, $8.0\%$ up one level, $12.9\%$ by two or more), reflecting the stricter centrality-aware rubric, which treats peripheral-but-matching items as a lower grade (Table~\ref{tab:relabel}); Table~\ref{tab:label-dist} gives the full label distribution. The sharp drop in the Exact grade ($501\!\to\!27$) follows directly from the guideline instruction to reserve grade~4 for rare, fully specified matches.

\subsection{Threats and Mitigations}
The principal threat is the centrality stringency gap between annotators. We mitigate it by (i) using a single canonical annotator who covers the whole corpus, so the labels are internally consistent; (ii) freezing the guidelines after a fully doubly annotated calibration batch; and (iii) reporting prevalence- and bias-robust coefficients alongside Cohen's $\kappa$ so the reliability is not overstated. The remaining limitation is that batches beyond the calibration set are single-annotated, so corpus-wide agreement is estimated from the $120$-pair calibration set, with the $13$ recurring cross-batch pairs as a consistency check, rather than measured on every item.

\subsection{Annotation Guidelines}
Annotators worked from a written rubric over a spreadsheet containing the query, product title, product description, the LLM-generated query- and product-attribute expansions, and the LLM ``final explanation''. For each row they performed four steps in order. (1) \textbf{Relevance} on a four-point scale: \textbf{4} (exact match: the title carries all defining details such as model, brand, or part number; annotators were told grade~4 should be \emph{rare} and not overused), \textbf{3} (close but not exact; the user very likely found the product), \textbf{2} (partial match; details missing but a close alternative), and \textbf{1} (little or nothing in common). When a query intent was unclear even after the description, it was flagged rather than graded. (2) \textbf{Centrality}, a binary judgment of whether the product is a central, typical, intent-satisfying result for the query (1) or a peripheral, atypical, or only indirectly relevant one (0); for example, for ``running shoes'' a standard pair is central while a specialised marathon shoe is an outlier, and for ``iphone'' a phone is central while a case is peripheral. The guideline notes that centrality is partly subjective. (3) \textbf{Explanation post-editing}: the LLM explanation was kept (\texttt{no edit}), lightly revised (\texttt{minor edit}), or rewritten (\texttt{major edit}). (4) A short free-text note recording what drove the decision (brand, type, demographic, material, price, condition). Annotators were asked to use consistent, plain terminology.

\subsection{Annotators and Cost}
Two annotators were hired for the project. A1 was compensated hourly at our institution's standard research-assistant scale (grade~3.6) for all project time, including annotation, guideline calibration, and meetings. A2 is an expert linguist holding a PhD in computational linguistics with prior data-annotation experience, employed as a postdoctoral researcher and compensated at the corresponding institutional salary scale. Both annotators co-designed the relevance, centrality, and post-edit rubrics and convened after each batch to review a sample of their work.

\section{Ablation: Human Annotation Features}
\label{sec:appendix-ablation}

To understand which human annotation signals drive the Q+P+H upper bound, we conduct a leave-one-out ablation over the three human-provided features: the post-edited relevance explanation (\textit{post-edit}), the free-text annotator note (\textit{comments}), and the centrality score (\textit{centrality}). Each configuration is retrained from scratch using the same protocol as the Q+P+H baseline and evaluated on the same 75-query test set. Results are shown in Table~\ref{tab:human-ablation}.

\begin{table}[ht]
\centering
\small
\begin{tabular}{lcc}
\toprule
\textbf{Configuration} & \textbf{nDCG@10} & $\boldsymbol{\Delta}$ \\
\midrule
All human annotations   & \textbf{0.9382}  & ---      \\
w/o \textit{centrality} & 0.9372           & $-0.0010$ \\
w/o \textit{comments}   & 0.8742           & $-0.0640$ \\
w/o \textit{post-edit}  & 0.8607           & $-0.0775$ \\
\bottomrule
\end{tabular}
\caption{Leave-one-out ablation of human annotation features on the Q+P+H
re-ranker. Each model is retrained from scratch; $\Delta$ is relative to the full human-annotation baseline.}
\label{tab:human-ablation}
\end{table}

The post-edited relevance explanation is the single most valuable human signal ($-7.75$~pp when removed), followed closely by the free-text comments ($-6.40$~pp). Both text fields encode reasoning that the LLM-generated explanations do not capture: annotators use the post-edit field to correct or sharpen the LLM's relevance assessment, and the comments field to record subtle distinctions (partial compatibility, demographic fit) that affect the final grade. The centrality score, by contrast, contributes negligibly ($-0.10$~pp): as a single scalar, it carries little information beyond what the relevance label already encodes. This finding has a direct practical implication: annotators can safely omit the centrality score with no measurable loss, while the two text fields should be retained whenever human annotation effort is budgeted.

\section{Qualitative Analysis}
\label{sec:appendix-qual}

\noindent\begin{minipage}{\textwidth}
\centering
\small
\setlength{\tabcolsep}{6pt}
\renewcommand{\arraystretch}{1.15}
\begin{tabular}{lcccc}
\toprule
Query & Q+P nDCG & Synth nDCG & $\Delta$ & Pattern \\
\midrule
\texttt{``nike shoes''} & 0.5157 & 0.9988 & +0.4831 & Brand-specific, low baseline \\
\texttt{``golf shoes''} & 0.5421 & 0.9857 & +0.4436 & Category-specific, low baseline \\
\texttt{``anime necklace''} & 0.5995 & 0.9620 & +0.3625 & Niche category \\
\texttt{``memory foam mattress topper''} & 0.6066 & 0.9299 & +0.3233 & Multi-word specific \\
\texttt{``minecraft diamond sword''} & 0.6947 & 0.9123 & +0.2176 & Niche product (gaming) \\
\bottomrule
\end{tabular}
\captionof{table}{Top improvements (queries where synthetic helps most).}\label{tab:top-improve-synth}
\end{minipage}

\vspace{0.75em}

\noindent\begin{minipage}{\textwidth}
\centering
\small
\setlength{\tabcolsep}{6pt}
\renewcommand{\arraystretch}{1.15}
\begin{tabular}{lcccc}
\toprule
Query & Q+P nDCG & Synth nDCG & $\Delta$ & Pattern \\
\midrule
\texttt{``light bulb''} & 1.0000 & 0.9676 & -0.0324 & Already perfect \\
\texttt{``levis''} & 0.9698 & 0.9530 & -0.0169 & High baseline \\
\texttt{``memory foam pillow''} & 0.9922 & 0.9785 & -0.0137 & High baseline \\
\texttt{``polaroid camera''} & 0.9931 & 0.9878 & -0.0054 & High baseline \\
\texttt{``electric griddle''} & 0.9926 & 0.9878 & -0.0048 & High baseline \\
\bottomrule
\end{tabular}
\captionof{table}{Top degradation (queries where synthetic hurts).}\label{tab:top-degrade-synth}
\end{minipage}

\section{Synthetic Generation of Human Annotated Features}
\label{sec:appendix-synth-ann}

\vspace{0.75em}

{\small
\centering
\begin{minipage}{\columnwidth}
\begin{lstlisting}[
    basicstyle=\ttfamily\small,
    breaklines=true,
    frame=single,
    aboveskip=0pt,
    belowskip=0pt,
    xleftmargin=0.15cm,
    xrightmargin=0.15cm,
    language=json
]
{
  "query": "memory foam mattress topper",
  "product_title": "Spa Sensations 4'' Memory Foam Mattress Topper, Multiple Sizes",
  "median_relevance": 4,

  "synthetic_annotations_qwen3.5_27b": {
    "relevance_score": 2,
    "centrality_score": 1,
    "comments": "The product is moderately relevant as it matches the query's classification type and user intent. However, the specific attribute '4''' (thickness) is not mentioned in the query, and the brand 'Spa Sensations' is not specified. While the core product matches, the lack of explicit alignment on specific dimensions or brand preferences prevents a higher score, fitting the 'Moderately relevant' category.",
    "post_editing_column": "Major edit"
  },

  "human_annotations": {
    "relevance_score": 2,
    "centrality_score": 0,
    "comments": "The product title is relevant to the query due to exact matches in classification type, user intent, and implied demographic, but lacks match in primary attribute focus and price indicator",
    "post_editing_column": "Minor edit"
  }
}
\end{lstlisting}
\end{minipage}
\captionof{figure}{Comparison of synthetic annotations (Qwen3.5-27B) versus human annotations for a query-product pair. Both annotators assigned a relevance score of $2$, but differed in centrality assessment ($1$ vs $0$) and post-editing comments. The synthetic model provides more detailed reasoning about thickness and brand specificity, while the human annotator focuses on classification and demographic alignment.}
\label{fig:synthetic-vs-human-annotation}
\par}

\vspace{0.75em}

\subsection{Synthetic--Human Agreement}
\label{subsec:synthetic-human-agreement}

To quantify how faithfully the synthetic annotations approximate the human-provided signals, we compare Qwen3.5-27B outputs against the corresponding human annotations on the same query-product pairs. Because relevance is ordinal, we report exact agreement, mean absolute error, and quadratic-weighted Cohen's $\kappa$; for binary centrality, we report exact agreement and Cohen's $\kappa$. These statistics measure annotation fidelity directly and complement the downstream ranking evaluation.

\begin{table}[ht]
\centering
\small
\begin{tabular}{llc}
\toprule
Signal & Agreement metric & Qwen3.5-27B vs. Human \\
\midrule
Relevance & Exact agreement & $36.1$\% \\
          & Mean absolute error & $0.84$ \\
          & Quadratic-weighted $\kappa$ & $0.09$ \\
\midrule
Centrality & Exact agreement & $60.7$\% \\
           & Cohen's $\kappa$ & $0.22$ \\
\bottomrule
\end{tabular}
\caption{Agreement between synthetic Qwen3.5-27B annotations and human annotations on the shared evaluation pairs. The low $\kappa$ for relevance ($0.09$) reflects a genuine disagreement -- the LLM has a strong ``Highly Relevant'' bias (never predicts scores $0-1$), while humans assign those scores to $\approx30$\% of pairs.}
\label{tab:synthetic-human-agreement}
\end{table}

\end{document}